\documentstyle[epsfig]{l-aa}
\def\infig#1#2#3{\epsfxsize=#3cm \centering{\mbox{\epsfbox{#2}}}\vspace{-0.4cm}}
 
\begin{document}

\thesaurus{06         
          (08.09.2;   
           08.02.2;   
	   08.06.3;   
	   03.20.4;   
           03.20.7)   
             }

\title{Physical parameters of the Algol system TZ~Eridani from
        simultaneous analysis of Geneva 7-colour light curves 
       \thanks{Based on observations made at the European Southern Observatory (La Silla, Chile)
               and at the Mount Laguna Observatory (USA)}
\thanks{Table~1 is only available in electronic form at the CDS via anonymous ftp to cdsarc.u.strasbg.fr
(130.79.128.5) or via http://cdsweb.u-strasbg.fr/Abstract.html}}
\author{F. Barblan \inst{1} \and P. Bartholdi \inst{1} \and P. North \inst{2} \and G. Burki \inst{1} \and E.C. Olson \inst{3}}
\offprints{F. Barblan}
\institute{Geneva Observatory, CH-1290 Sauverny, Switzerland 
   \and     Institut d'Astronomie de l'Universit\'e de Lausanne, CH-1290 Chavannes-des-Bois, Switzerland
   \and Astronomy Department, University of Illinois, 1002 W. Green St., Urbana, Illinois 61801, USA}

\date{Received April 1, 1998; accepted }
\maketitle
\markboth{F. Barblan et al.: Physical parameters of the Algol system TZ Eridani}{F. Barblan et al.: Physical parameters of the Algol system TZ Eridani}
\begin{abstract}

Light curves of the semi-detached eclipsing binary system TZ~Eridani in the Geneva 7-colour photometric
system were analysed using the Wilson-Devinney programme. The physical and orbital parameters have been determined
through a self-consistent simultaneous solution of the seven light curves and of the radial velocity 
curves of both components. The following absolute elements of the components are, for the primary (mass gainer), 
$M_1 = 1.97 \pm 0.06$ M$_{\odot}$, $R_1 = 1.69 \pm 0.03$ R$_{\odot}$, $M_{\rm bol_1} = 2.36 \pm 0.09$, 
$T_{\rm eff_1} = 7770 \pm 100$ K, and for the secondary (mass loser), $M_2 = 0.37 \pm 0.01$ M$_{\odot}$, 
$R_2 = 2.60 \pm 0.04$ R$_{\odot}$, $M_{\rm bol_2} = 3.74 \pm 0.13$, $T_{\rm eff_2} = 4570 \pm 100$ K. The semi-major axis $A$ 
of the relative orbit is $10.57 \pm 0.16$ R$_{\odot}$. 
The regular increase of the period is described. The spectral type of the components are A5/6~V (primary) and K0/1~III. 
The secondary has exhibited a long-term luminosity increase of about 0.06 in $V$ between Dec~1983 and Dec~1996. The distance
to TZ~Eri is evaluated to 270 $\pm$ 12 pc.

\end{abstract}

\section{Introduction}

TZ~Eridani (BD -6$\degr$880) is an Algol-type EA eclipsing binary of short period, $P = 2.606$ days.
The total primary eclipse is 2.5 mag deep in the $V$ band. The estimation of the spectral type of
the two components is F (Cannon, 1934), F8 (Brancewicz \& Dvorak, 1980) for the primary, and
K0IV (Kaitchuck \& Park, 1988), K0 (Yoon et al., 1994) for the secondary. However, as shown in 
this paper, the most probable spectral types are A5/6~V for the primary and K0/1~III for the
secondary.

A first estimate of the physical parameters of the components is given by Brancewicz \& Dvorak (1980), 
in their ``Catalogue of Parameters for Eclipsing Binaries'', in particular $R_{1}= 1.51$ R$_{\odot}$ 
(primary component) and $R_{2}= 1.92$ R$_{\odot}$. Recall that these values have not been obtained by 
a simultaneous treatment of the complete light and radial velocity curves. It is the aim of this 
paper to perform such a new analysis, by using the method and the computer programme of Wilson \& Devinney
(1971).

No complete light or radial velocity curves have been yet published for TZ Eri. For that reason, 
this star was measured intensively in the 7-colour Geneva photometric system (Golay, 1980; 
Rufener, 1988) using the Swiss telescope at La~Silla (European Southern Observatory, Chile) equipped 
with the two-channel aperture photometer P7 (Burnet \& Rufener, 1979). Moreover, the radial velocity
curve of the primary component has been determined with the spectrovelocimeter \textsc{Coravel} installed
on the 1.54~m Danish telescope at La~Silla, and with the Illinois Cassegrain (``white'') spectrograph attached to the
1~m Illinois reflector at Mount Laguna Observatory. In addition, one spectrum of the secondary 
component has been obtained by using the NTT 3.5~m telescope at La~Silla, allowing the determination of the 
masses of the two components of TZ~Eri.

TZ~Eri was discovered to have a disk (Kaitchuck \& Honeycutt, 1982) on the basis of emission in 
H$_{\beta}$ and H$_{\gamma}$ lines observed during the eclipse. Kaitchuck \& Park (1988) showed that 
this disk belongs to the class of the transient accretion disks which are produced by a collision 
of the gas stream with the mass-gaining star (primary component). The radius of the disk was 
measured by the duration of the presence of H$_{\beta}$ line in emission after the start of the 
primary eclipse (second contact) or before the end of the same eclipse (third contact). The disk 
extension is variable from one eclipse to another, and in most cases between the trailing and 
leading sides. During the 12 eclipses studied, the disk extension $r/R_{1}$ varied from 1.0 to 1.64.

In this paper, the variation of the period of TZ~Eri will be analysed, the variability of the 
components will be examined, and the physical parameters of the two components will be determined 
from the analysis of the light and radial velocity curves.

\section{Period}

The orbital period listed in the GCVS (Kholopov, 1985) is $P = 2.6060653$ days.
However, we are dealing here with a semi-detached system which, by definition,
has a variable period due to mass transfer. During most photometric observations
done in 1984-1987 by Dr. Zdenek Kviz, the period remained fairly constant, since it was
possible to obtain a well-defined lightcurve with $P=2.6061082$ days. This
value was obtained using the reciprocal $\theta_1$ test of Renson (1978) 
-- see also Manfroid et al. (1991) -- which had been defined above all for Ap
stars (which vary with small amplitudes) but proved very efficient for
a precise determination of the period of eclipsing binaries.
The ephemeris we have adopted is~:

\begin{eqnarray}
\mbox{HJD}(\mbox{Min I})& = & ~~~(2\,446\,109.6922 ~~ \pm {\it 0.0010}) \\ 
                        &   & + (2.6061082 ~~ \pm {\it 0.0000020}) \times E \nonumber
\end{eqnarray}
The orbital period has varied, as shown by the times of minima registered
over decades by amateur astronomers. The $O\!-\!C$ values so obtained, and
published in the BSAG Bulletin (e.g. Locher 1997) are shown in Figure 1
together with a parabolic fit which gives~:

\begin{eqnarray}
\mbox{O-C} & =  & - (0.0050 ~~ \pm {\it 0.0014}) \\ 
           &    & - (1.008 \pm {\it 0.366}) \times 10^{-6} (\mbox{HJD}-t_0) \nonumber \\
           &    & + (2.433 \pm {\it 0.143}  \times 10^{-9} (\mbox{HJD}-t_0)^2  \nonumber 
\end{eqnarray}
where $t_0$ = 2\,446\,109.6922 . The typical error on the epochs of primary minima 
in Fig.~1 is a few minutes (0.002 - 0.005 d.). Therefore
the fit only represents a mean trend, upon which are superposed sudden period
changes that cannot be accounted for by measurement errors (see especially the
very steep rise just before JD 2\,450\,000).
\begin{figure}
\infig{8.8}{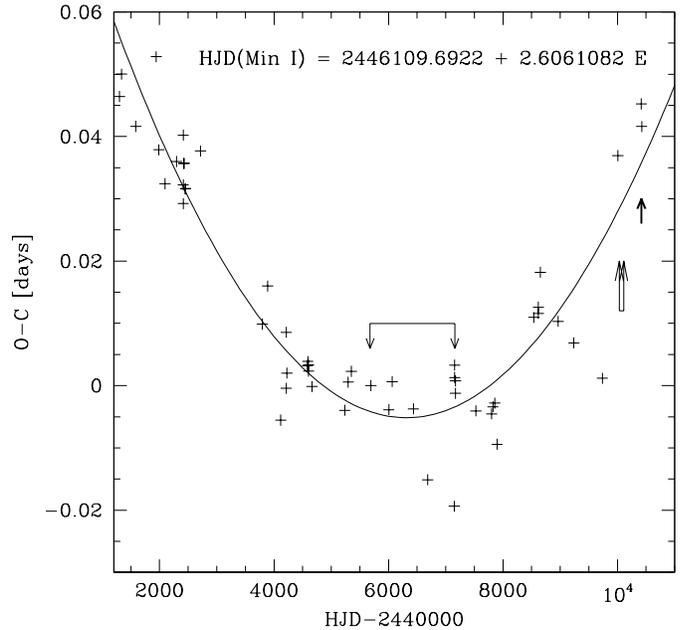}{8.8}
\caption[]{O-C diagram of TZ Eri from observations made by amateurs and
published in the BSAG Bulletin (two points are defined by Geneva photometry).
A parabolic fit is superposed, showing the
regular increase in the period. Short arrows define the intervals of photometric
measurements; long arrows define the interval of radial-velocity observations.}
\end{figure}

Although the period changes are interesting by themselves, they are rather
a nuisance in our context, because the radial-velocity measurements, which were
made recently, have to be put in phase with the photometric ones, which are
much older. For this reason we have not used the above formula for $O\!-\!C$, 
but we have simply used additional photometric measurements kindly made by Marc 
K\"unzli in November 1996 with the same equipment. He has made 36 new 
multicolour measurements, several of which during the primary minimum. Using 
the code EBOP16 (Etzel, 1989) and the adopted ephemeris, we adjusted the 
$\Delta\theta$  parameter (phase correction for the position of the primary 
minimum) for these recent data, as well as for the old data alone. The 
difference is~:
 
\begin{equation}
\Delta\phi = -0.01734\pm 0.00020
\end{equation}
and represents the phase correction to be 
applied to the 1996 data, to fit them into the adopted ephemeris. Although the 
$V_r$ observations have been made one year earlier than the new photometric 
ones, we neglect the slight period change that may have occurred in between,
compared to the change that has taken place between the old photometric
measurements and the $V_r$ ones. Therefore, the above phase shift was applied
as such to the $V_r$ data.

\section{Photometric data and variability of the components}

Geneva 7-colour photometric measurements of TZ~Eridani were obtained
from Dec 10, 1983 to Dec 10, 1996, using the Swiss
70~cm telescope at the European Southern Observatory (ESO), La Silla, Chile.
During this period, 393 measurements of weight $q \geq 1$ have been obtained
(see Rufener, 1988, for the definition of the weight q). These data are listed 
in Table~1, together with the 36 additional photometric measurements obtained 
in Nov-Dec 1996 (see Section~2).

The magnitudes in each of the seven filters are obtained from the visual 
magnitude $V$ and the six colour indices in the following manner : 

\begin{equation}
i = V - [V - B] + [i - B]
\end{equation}
with i representing one of the seven filters $U$, $B$, $V$, $B_{1}$, $B_{2}$,
$V_{1}$, G. Remember that the Geneva $[U - B]$ and $[B - V]$ indices are not
normalized to zero for an A0V star as it is the case for the Johnson UBV 
indices. 
It is possible to calculate the magnitude of the primary (mass gainer) by subtracting 
the flux of the secondary (mass loser), at the bottom of the primary eclipse, from
the flux of the both components measured together (outside the eclipses). 
This calculation has been made for each of the seven Geneva
magnitudes. In order to minimize the effects of a possible long-term variability 
(see the end of this Section), only the data obtained in December 1983 and 
January 1984 have been used. The results are given in Table~2.

It is interesting to compare
the observed uncertainties with the mean precision of the measurements
made in Geneva photometry. Rufener (1988, Fig.~2) has shown the shape
of the mean relation $\sigma_{V}$ vs. $V$ obtained for the non-variable
stars, in particular the progressive increase of $\sigma_{V}$ with
increasing $V$, for stars fainter than $V \simeq 9$. The same relation can 
be applied to the seven Geneva magnitudes. Fig.~2 of this paper shows
a new calculation of this relation, based on the up to date version of our
photometric database. On the same figure are plotted the observed values for TZ~Eri given
in Table~2. A correction has been applied to the uncertainties of the
magnitudes for the secondary, because the measurements obtained during the totality
of the primary eclipse had shorter integration time than the other ones (4 minutes
instead of 12 minutes). The conclusion is that the uncertainties on the measurements
of TZ~Eri are in agreement with the expected precision. Thus, the components do not
exhibit a short-term variability, i.e on a time-scale shorter than about 20 orbital 
periods. 

\setcounter{table}{1}
\begin{table}[hbt]
\caption{The seven Geneva apparent magnitudes of TZ~Eri (measures of December 1983 and
          January 1984).} 
\begin{center}
\begin{tabular}{lccc}  
\hline\rule{0pt}{2.3ex}%
Mag.	     & TZ~Eri A+B & Secondary & Primary \\
             & Observed	& Observed      & Calculated     \\[0.3ex]
\hline\rule{0pt}{2.3ex}%
U  & $ 10.540 \pm 0.013$ & $ 14.87  \pm 0.17  $ & $ 10.560 \pm 0.016 $ \\
B1 & $  9.987 \pm 0.009$ & $ 13.980 \pm 0.050 $ & $ 10.015 \pm 0.011 $ \\
B  & $  9.011 \pm 0.009$ & $ 12.584 \pm 0.031 $ & $  9.052 \pm 0.010 $ \\
B2 & $ 10.409 \pm 0.007$ & $ 13.670 \pm 0.035 $ & $ 10.464 \pm 0.009 $ \\
V1 & $ 10.299 \pm 0.011$ & $ 12.809 \pm 0.021 $ & $ 10.406 \pm 0.014 $ \\
V  & $  9.576 \pm 0.010$ & $ 12.067 \pm 0.019 $ & $  9.691 \pm 0.013 $ \\
G  & $ 10.679 \pm 0.016$ & $ 12.960 \pm 0.018 $ & $ 10.821 \pm 0.021 $ \\
\hline	 	        	 	  
\end{tabular}
\end{center}
\end{table}

\begin{figure}
\infig{8.8}{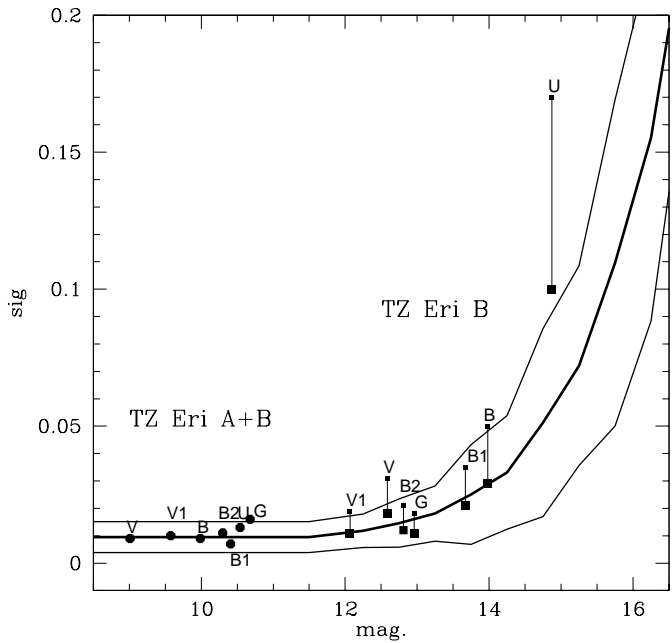}{8.8}
       \caption[ ]{Variation of the mean precision $\sigma$ 
       with the magnitude, in the case of Geneva photometric measurements. 
       The solid thick line refers to the mean value and the solid
       thin lines to the 1 s.d. level. Dots and small squares concern respectively the seven
       magnitudes of TZ~Eri~A+B outside eclipses and of the secondary (during the totality of the 
       primary eclipse). Big squares represent the estimated values of $\sigma$ for the secondary 
       which ought to have resulted from a ``normal'' integration time of the measurements, 
       i.e. 12 minutes (normal) instead of 4 minutes (during the primary eclipse).} 
       \end{figure}

The long-term photometric behaviour of both components has been analysed by comparing our 
photometric data obtained at 4 epochs, corresponding to the intensive monitoring of the
eclipses~: Dec 1983 to Jan 1984, Jan 1985, Nov 1987 and Nov-Dec 1996. Table~3 gives the mean 
values of $V$, $[B-V]$ and $[U-B]$ at each of these epochs for the both components. It appears 
that~: 

\begin{enumerate}
\item
   The secondary exhibited a long-term luminosity increase (0.06 in $V$) between Dec 1983
   and Dec 1996. The colour variations are large, especially in $[U-B]$, but not significant due
   to the large standard deviation.
\item
   The primary did not show any long-term variation in magnitude or in colours.
\end{enumerate}

The observed variations of TZ~Eri secondary are similar to those studied by Olson \& Etzel (1993)
in six cool subgiant secondaries of totally eclipsing Algol systems. They noted that the fluctuations 
increase with decreasing orbital period, or with increasing rotational velocity, suggesting that
rotationally induced magnetic activity could be the origin of these brightness variations.

In order to minimize the effects of the long-term variation of the secondary luminosity, only
the data obtained before HJD 2\,447\,200 (Feb 1988) have been used for the eclipse analysis
(see Section~6).

\begin{table*}[hbt]
\caption{Long-term behaviour of the components of TZ~Eri in $V$ magnitude and Geneva colours
$[B-V]$ and $[U-B]$. Only the variation of the secondary in $V$ is significant.} 
\begin{center}
\begin{tabular}{|c|rrr|rrr|}  
\hline \rule{0pt}{2.3ex}
	 & \multicolumn{3}{c|}{Secondary (mass loser)} & \multicolumn{3}{c|}{Primary (mass gainer)} \\
         &    $V$   & $[B-V]$ & $[U-B]$ &   $V$   &  $[B-V]$ & $[U-B]$ \\[0.3ex]
\hline \rule{0pt}{2.3ex}
Dec 1983 & 12.067 & 0.523 & 2.281 & 9.691 & -0.639 & 1.508 \\
	 & $\pm 0.019$ & $\pm 0.046$ & $\pm 0.175$ & $\pm 0.013$ & $\pm 0.023$ & $\pm 0.026$ \\
Jan 1985 & 12.045 & 0.466 & 2.184 & 9.691 & -0.639 & 1.561 \\
Nov 1987 & 12.035 & 0.453 & 1.990 & 9.692 & -0.637 & 1.561 \\
Dec 1996 & 12.007 & 0.477 & 2.189 & 9.695 & -0.635 & 1.547 \\
\hline
\end{tabular}
\end{center}
\end{table*}

\section{Classification of the components and interstellar extinction}

A photometric classification of the components of TZ~Eri can be obtained 
because the primary eclipse is total, by using various calibrations
based on the colour indices and reddening free parameters of the Geneva
photometric system. 

From the magnitudes in Table~2, the various Geneva colour indices and 
reddening free parameters $d$, $\Delta$ and $g$ (see Golay, 1980) have
been calculated. Then, the technique of the {\it photometric boxes} (Golay et al.,
1969; Nicolet, 1981a) has been applied to determine the intrinsic characteristics
of the primary component. Recall that this technique is specific to the Geneva photometric system 
and depends on the homogeneity and precision of its measurements. The assumption
in this technique is that the properties of the stars lying within a specified
photometric neighbourhood (see Nicolet, 1994) can be equated with each other,
provided that the radius of the photometric box is small enough. We applied the
technique to the parameters $d$, $\Delta$ and $g$ of the primary and found in the
entire photometric database 40 twin stars, i.e. stars having very similar values 
of the 3 parameters, the radius of the photometric box having been chosen equal
to 0.015 mag. It is assumed that these stars are intrinsically similar to the primary of TZ~Eri.
The same method was applied to the classification of the components of the
eclipsing RS~CVn-type system RZ~Eridani (Burki et al., 1992).

It is especially noteworthy that 6 of these twin stars belong to open clusters~:
$\alpha$Per, Hyades, Praesepe, Pleiades (2 stars) and NGC~6405. This allowed to
determine unanbiguously the intrinsic colours of these stars, since these clusters
have well determined distances and interstellar extinctions. By using the values
given by Nicolet (1981b), we derived the intrinsic colours of the twin stars, and
thus also of the primary of TZ~Eri, in particular $[B2-V1]_0 = 0.011 \pm 0.016$. According to the
relations between spectral types and intrinsic Geneva colours or parameters by
Hauck (1994) and to the calibration of Geneva photometry based on Kurucz's 
atmosphere models by K\"{u}nzli et al. (1997), the estimated spectral type of
the primary is A5/6~V, its effective temperature is $7770 \pm 100$~ K, its gravity
is $\log g = 4.40 \pm 0.07$ and its metallicity is $[M/H] = 0.13 \pm 0.09$. On the other 
hand, the colour excess is $E[B2-V1] = 0.047 \pm 0.017$.

For the secondary, we have calculated the intrinsic colours by using the measured values 
(see Table~1) and the colour excesses obtained for the primary. We obtained in
particular $[B2-V1]_0 = 0.756$, an estimated spectral type of K0/1~III and an 
effective temperature of $4520$~K.

\section{Radial velocity curves}

The primary's radial velocity curve has been measured with the \textsc{Coravel} scanner
(Baranne et al. 1979) in December-January 1995-1996. The cross-correlation dip is rather wide and
shallow due to the fast axial rotation, so that the scatter of the residuals
around the fitted curve is rather large (several km\,s$^{-1}$). The secondary
component remained invisible to \textsc{Coravel}, due to its low luminosity in the blue.
About one year later, 14 new observations have been done by E.C.~Olson using the
Illinois Cassegrain (``white'') spectrograph~: 38~cm focal length camera, 831 line mm$^{-1}$ grating, 
Texas Instruments 800$\times$800 CCD detector cooled with liquid nitrogen; the CCD was operated by
a Photometrics Ltd. controller; spectral resolution in the first order is 0.04\AA per pixel; 
Reductions were done with IRAF.

There has been a problem in fixing the zero-point of the radial velocities obtained
with this instrument, so that we fitted an orbit separately to the corresponding
data to obtain the apparent systemic velocity. Then we applied to these $V_r$
values a uniform shift equal to the difference between this apparent systemic
velocity and that obtained with \textsc{Coravel} measurements alone. Indeed, we are
confident in the \textsc{Coravel} $V_r$ scale, because several standard stars have
been observed each night and the small instrumental drifts ($\pm 1$~km\,s$^{-1}$
at most) are well controlled. Therefore, the uncertainties quoted in Table 5
may be slightly optimistic, because they refer to an orbital solution which
assumed a perfect correction to Olson's radial velocities.

In order to see the secondary star's spectrum and to obtain the mass ratio of the
components, we asked Dr. Didier Raboud to observe TZ Eri in the vicinity of
a quadrature with the NTT telescope at ESO. He could indeed take one spectrum,
with an exposure time of 10~min,
on 11th November 1995, using the EMMI spectrograph in the REMD mode, with
Grism \#5, Grating \#10 and a slit measuring $1"\times 6"$; in this
configuration, the resolving power is $R=28000$ and the wavelengths range
between 4013 and 6606 \AA. The detector was CCD \#36 (ESO numerotation), a thin,
back-illuminated Tektronix TK2048EB chip with $24\times 24\,\mu{\rm m}$ pixels.
The spectrum has been reduced at Geneva Observatory by Mr. Michel Studer, using
the \textsc{Tacos} software developed by Dr. Didier Queloz for the 
\textsc{Elodie} spectrograph at Observatoire de Haute-Provence.
The radial velocities were obtained by cross-correlation between the observed
spectrum and a binary mask optimized for F0-type stars, which yielded two dips,
one for each component. Thanks to the long wavelength interval extending well
into the red, the cool companion is easily seen in the correlation function.
The dips are only 3.6\% and 1.1\% deep for the primary and secondary 
respectively, but the S/N ratio of the correlation function is better than 500.
A K0-type mask was tried as well and yielded the same result, but with no
improvement.

The journal of the radial velocity observations is given in Table 4 and the 
radial velocity curve is shown in Figure~3. Additional spectra
would of course be welcome to complete the secondary's $V_r$ curve, but the
single point we have suffices to constrain the mass ratio to a precision of
about 7\% ($q$ = 0.193 $\pm$ 0.013). The orbital elements are given in Table~5.

\begin{table}
\caption{Journal of the radial velocity observations of TZ Eri. The phases
are computed from the ephemeris given by Equation~1 and corrected for period
change according to Equation~3.}
\begin{center}
\begin{tabular}{ccrlcc}
\hline \noalign{\smallskip}
HJD&comp.&\multicolumn{2}{c}{$V_r \pm \sigma$}&Instrument&phase\\
$\null-2\,400\,000$&&\multicolumn{2}{c}{[km s$^{-1}$]}&or observer&\\
\noalign{\smallskip} \hline\noalign{\smallskip}
50033.785 &A&  18.08& 2.   &\textsc{Emmi} & 0.712 \\
          &B&-173.40& 3.   &       &       \\
50039.606 &A&   0.67& 2.87 &\textsc{Coravel}& 0.945 \\
50042.599 &A& -27.19& 3.08 &       & 0.094 \\
50083.578 &A&  10.32& 2.50 &       & 0.818 \\
50084.542 &A& -42.17& 2.55 &       & 0.188 \\
50084.708 &A& -48.40& 2.89 &       & 0.252 \\
50085.544 &A&  -1.72& 1.87 &       & 0.572 \\
50085.707 &A&  11.17& 2.30 &       & 0.635 \\
50086.550 &A&   3.38& 1.56 &       & 0.958 \\
50087.540 &A& -41.39& 2.17 &       & 0.338 \\
50087.684 &A& -38.53& 2.46 &       & 0.394 \\
50088.551 &A&  17.60& 1.96 &       & 0.726 \\
50088.709 &A&  15.04& 2.63 &       & 0.787 \\
50097.630 &A& -42.39& 2.23 &       & 0.210 \\
50098.606 &A&   3.24& 2.37 &       & 0.584 \\
50110.324 &A& -29.78& 3.29 &       & 0.081 \\
50407.863 &A& -47.05& 2.5  &\textsc{E.C.~Olson}& 0.251 \\
50407.880 &A& -44.75& 2.5  &       & 0.257 \\
50407.897 &A& -46.65& 2.5  &       & 0.264 \\
50408.730 &A&   3.45& 2.0  &       & 0.583 \\
50411.717 &A&  18.55& 2.0  &       & 0.729 \\
50411.745 &A&  20.75& 2.0  &       & 0.740 \\
50411.872 &A&  16.35& 2.0  &       & 0.789 \\
50412.697 &A& -34.05& 2.0  &       & 0.106 \\
50412.719 &A& -35.95& 2.0  &       & 0.114 \\
50412.853 &A& -40.95& 2.0  &       & 0.165 \\
50413.735 &A& -15.05& 2.5  &       & 0.504 \\
50413.931 &A&   3.15& 1.5  &       & 0.579 \\
50415.700 &A& -39.65& 2.5  &       & 0.258 \\
50415.725 &A& -40.15& 2.5  &       & 0.267 \\
\noalign{\smallskip} \hline
\end{tabular}
\end{center}
\end{table}
\begin{figure}
\infig{8.8}{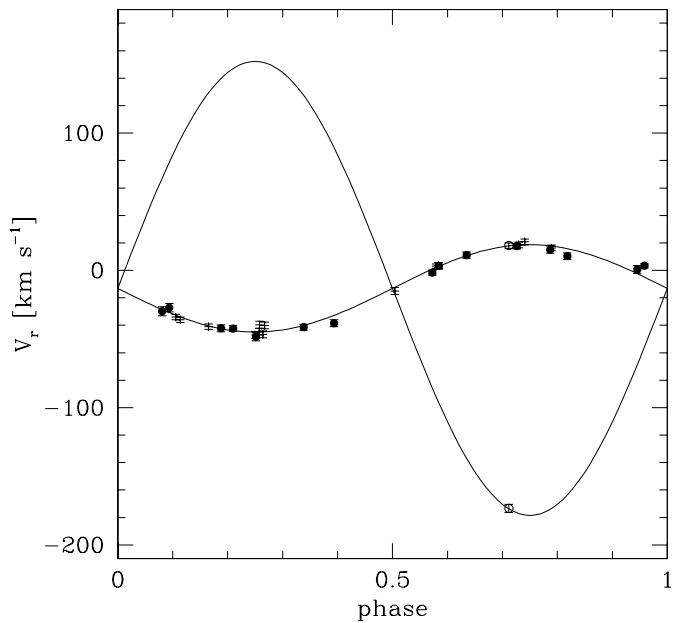}{8.8}
\caption[]{Radial velocity curve of TZ Eridani. The black dots (\textsc{Coravel}
observations) and the plus signs (Olson's observations) represent
the primary component, while the open dot represents the secondary. The phases
are those of Table~4.}
\end{figure}

\begin{table*}
\caption{Orbital elements of the binary. For each component, the
second line gives the estimated standard deviations of the
parameters. A null uncertainty means that the corresponding parameter
has been fixed before the convergence. The period has been fixed to the
value it had in the middle of the time spanned by the $V_r$ observations.}
\begin{center}
\begin{tabular}{|r|r|r|r|r|r|r|r|r|r|r|}
\hline\rule{0pt}{3.5ex}%
Star name & \multicolumn{1}{|c|}{$P$} & 
\multicolumn{1}{|c|}{$T_\circ$ [HJD} & \multicolumn{1}{|c|}{$e$} & 
\multicolumn{1}{|c|}{$V_\circ$} & \multicolumn{1}{|c|}{$\omega_1$} & 
\multicolumn{1}{|c|}{$K_{1,2}$} & 
\multicolumn{1}{|c|}{M$_{1,2} \sin^3 i$} & 
\multicolumn{1}{|c|}{a$_{1,2}\sin i$} & 
\multicolumn{1}{|c|}{N} & \multicolumn{1}{|c|}{(O-C)} \\
 & \multicolumn{1}{|c|}{[days]} & \multicolumn{1}{|c|}{-2\,450\,000]} &  & 
\multicolumn{1}{|c|}{[km s$^{-1}$]} & \multicolumn{1}{|c|}{[$^\circ$]} & 
\multicolumn{1}{|c|}{[km s$^{-1}$]}  & 
\multicolumn{1}{|c|}{[M$_\odot$]} & \multicolumn{1}{|c|}{[$10^{6}$ km]} &  & 
\multicolumn{1}{|c|}{[km s$^{-1}$]} \\[2.ex]
\hline\rule{0pt}{3.5ex}%
TZ Eri A    &     2.606132   &   83.4009  & 0.000  & -13.10  & 
               0.0  & 31.93  &  1.741  &    1.144  &  30  & 3.04\\  
            &     0.0000000  &    0.0000  & 0.000  &   0.55  & 
               0.0  &  0.70  &  0.087  &    0.025  &      &    \\  
         &  &  &  &  &  &  &  &  &  & \\
TZ Eri B    &  &  &  &  &  & 165.32  &  0.336  &    5.92  &   1  & \\  
         &  &  &  &  &  &  3.73  &  0.017  &    0.13  &   & \\  \hline
\end{tabular}
\end{center}
\end{table*}

\begin{table}[hbt]
\caption{Characteristics of the 7 Geneva passbands (Rufener \& Nicolet, 1988). $\lambda_{o}$ is
the mean wavelength and $\mu$ is the second-order moment (approximately half the passband width).}
\begin{center}
\begin{tabular}{lcc}  
\hline\rule{0pt}{3.5ex}%
Passband	     & $\lambda_{o}$[\AA] & $\mu$ [\AA]\\[2.ex]
\hline\rule{0pt}{3.5ex}%
U  & 3464 & 159 \\
B1 & 4015 & 188 \\
B  & 4227 & 282 \\
B2 & 4476 & 163 \\
V1 & 5395 & 202 \\
V  & 5488 & 296 \\
G  & 5807 & 200 \\
\hline	 	        	 	  
\end{tabular}
\end{center}
\end{table}

\begin{table}
\caption{Adopted values of the logarithmic limb-darkening parameters which were
kept fixed in the least-squares solution. For the primary, only the $y$ parameters
were fixed (the fitted $x$ parameters are listed in Table 8). For the secondary,
both $x$ and $y$ parameters had to be fixed.}
\begin{center}
\begin{tabular}{lrrr}
\hline\rule{0pt}{2.3ex}%
Passband & $y_1$ & $x_2$ & $y_2$ \\[0.3ex] 
\hline\rule{0pt}{2.3ex}%
U &0.243&0.940&-.515 \\
B1&0.303&0.863&-.240 \\
B &0.303&0.858&-.240 \\
B2&0.303&0.868&-.240 \\
V1&0.263&0.847&-.043 \\
V &0.263&0.825&-.043 \\
G &0.263&0.819&-.043 \\ \hline
\end{tabular}
\end{center}
\end{table}

\section{Rotation and Metallicity}

The rotational velocity of the primary deduced from the standard calibration
of the width of the correlation dip measured with \textsc{Coravel} (Benz \& Mayor, 1984), is
\begin{equation}
v\sin i = 38.1 \pm 2.2 km\,s^{-1}
\end{equation}
which is slightly faster than the synchronous rotation value, since with a radius
$R=1.70 R_\odot$ (see Section~6), the expected equatorial velocity of the primary 
is 33.0 km\,s$^{-1}$. It is not exceptional for Algol gainers to rotate faster than 
synchronously as shown by Koch et al. (1965). However, the difference being only 
$2.3 \sigma$, we cannot firmly conclude that a departure from synchronism exists in TZ~Eri.
On the other hand, $v\sin i$ may be overestimated, due to several factors. First, the
rotational velocity is so large that we hardly reach the continuum of the
correlation function, which can bias the fit. Second, the profile is fitted
by a gaussian, while the rotational profile itself does not have this shape,
even if it is convoluted by a gaussian instrumental profile (the calibration
was intended for slow rotators, with $v\sin i$ generally slower than $\sim$
20 kms$^{-1}$). Third, the
calibration has been devised above all for cool, solar-type stars (although
it does include a temperature term) and we apply it to a late A star, i.e.~at
the very border of its validity. Fourth, the $v\sin i$ determination is based on the 
hypothesis of a solar-type macroturbulence; this is probably not valid in a semi-detached
short period system and, thus the value given in (5) could be an overestimate.

We did not attempt to derive the projected rotational velocity of the secondary
component by using the correlation dip obtained with the NTT spectrum, because
a new calibration would have been needed which is beyond the scope of this
paper, and the correlation dip of the secondary is in any case very shallow. 

It is interesting to notice that the ``surface'' or equivalent width of the
\textsc{Coravel} autocorrelation dip of the primary is the same as for single late-A
stars. We have $W=2.31\pm 0.14$ km\,s$^{-1}$, while HD 2628, type
A7III and [Fe/H]=-0.02, has $W=2.41$ and HD 110379, type F0V and [Fe/H]=-0.07,
has $W=2.49$. This implies that the metallicity of TZ Eri is close to the solar value,
in agreement with the result obtained in Section~4 on the basis of the photometric
analysis.

\section{Photometric solution}

\begin{table*}[hbt!]
\caption[]{Adjusted parameters of the system TZ~Eri from the Wilson-Devinney programme. The parameters are~:\\
	 the temperatures $T_{1}$ and $T_{2}$ of the primary (hot) and secondary (cool) components,\\
	 the semi-major axis $A$ of the relative orbit,\\
	 the orbital inclination $i$, \\
	 the mass ratio $q$ (the uncertainty assumes exact lobe filling for the secondary),\\
	 the potential of the surface of the primary component $\Omega_{1}$ (for the units, see Section~7 and WD programme),\\
	 the potential of the surface of the secondary component $\Omega_{2}$ (for the units, see Section~7 and WD programme),\\
	 the exponent of the gravity darkening law $g_{2}$,\\
	 the normalized monochromatic luminosity in the seven Geneva passbands for the primary   $L_{1}/(L_{1}+L_{2})$,\\
	 the normalized monochromatic luminosity in the seven Geneva passbands for the secondary $L_{2}/(L_{1}+L_{2})$,\\
	 the center-to-limb darkening factors for the primary $x_{1}$ in the seven passbands.}
\begin{center}
\begin {tabular}{|l|ll|ll|ll|} 
\hline\rule{0pt}{3.5ex}%
Parameters                   &Values               &Uncertainties &Values               &Differences        &Values                 & Differences \\
                             &$(T_{1}=7770 K)$     &	          &$(T_{1}=7670 K)$     &{\it "7670--7770"} &$(T_{1} = 7870 K)$     & {\it "7870--7770"} \\[2.ex]
\hline\rule{0pt}{3.5ex}%
$T_{1}$ [K]                  &{\bf 7770}           &              &{\bf 7670}           &-100               &{\bf 7870}             &+100 \\[2.ex]

$T_{2}$ [K]                  &4563                 &$\pm$2        &4614                 &51                &4669                   &106 \\
$A$ [$R_{\odot}$]            &10.57                &$\pm$0.16     &10.59                &0.02              &10.56                  &-0.01 \\
$i$ [$\degr$]                &86.73                &$\pm$0.03     &86.77                &0.04              &86.71                  &-0.02 \\
$q=M_{2}/M_{1}$              &0.1865               &$\pm$0.0003   &0.1862               &-0.0003           &0.1867                 &0.0002 \\[2.ex]
				                                                                           
$\Omega_{1}$                 &6.46                 &$\pm$0.01     &6.45                 &-0.01              &6.49                  &0.03 \\
$\Omega_{2}$                 &2.20                 &              &2.20                 &0.0               &2.20                   &0.0 \\
$g_{2}$                      &0.322                &              &                     &                   &                       & \\[2.ex]
$L_{1}/(L_{1}+L_{2})$        &                     &              &                     &                   &                       &\\
\hspace{10mm} $U$            &0.9687               & $\pm$0.0010  &0.9692               &0.0005            &0.9692                 &0.0005 \\
\hspace{10mm} $B1$           &0.9644               & $\pm$0.0003  &0.9644               &0.0                &0.9644                 &0.0 \\
\hspace{10mm} $B$            &0.9481               & $\pm$0.0003  &0.9481               &0.0                &0.9481                 &0.0 \\
\hspace{10mm} $B2$           &0.9312               & $\pm$0.0003  &0.9313               &0.0001            &0.9313                 &0.0001 \\
\hspace{10mm} $V1$           &0.8763               & $\pm$0.0004  &0.8765               &0.0002            &0.8765                 &0.0002 \\
\hspace{10mm} $V$            &0.8676               & $\pm$0.0005  &0.8678               &0.0002            &0.8679                 &0.0003 \\
\hspace{10mm} $G$            &0.8420               & $\pm$0.0003  &0.8422               &0.0002            &0.8421                 &0.0001 \\[2.ex]
				                                                                           
$L_{2}/(L_{1}+L_{2})$        &                     &		  &			&		    &			    &\\
\hspace{10mm} $U$            &0.0313               & $\pm$0.0023  &0.0308               &-0.0005             &0.0308                 &-0.0005 \\
\hspace{10mm} $B1$           &0.0356               & $\pm$0.0006  &0.0356               &0.0                &0.0356                 &0.0 \\
\hspace{10mm} $B$            &0.0519               & $\pm$0.0005  &0.0519               &0.0                &0.0519                 &0.0 \\
\hspace{10mm} $B2$           &0.0688               & $\pm$0.0005  &0.0687               &-0.0001             &0.0687                 &-0.0001 \\
\hspace{10mm} $V1$           &0.1237               & $\pm$0.0006  &0.1235               &-0.0002             &0.1235                 &-0.0002 \\
\hspace{10mm} $V$            &0.1324               & $\pm$0.0006  &0.1322               &-0.0002             &0.1321                 &-0.0003 \\
\hspace{10mm} $G$            &0.1580               & $\pm$0.0004  &0.1578               &-0.0002             &0.1579                 &-0.0001 \\[2.ex]
\cline{4-7}				                                                                           
$x_{1}$                      &			   &		  & \multicolumn{4}{c|}{~}   \\
\hspace{10mm} ${U}$          &0.655                &$\pm$0.037    & \multicolumn{4}{c|}{~}   \\
\hspace{10mm} $B1$           &0.770                &$\pm$0.009    & \multicolumn{4}{c|}{Determinant of normal equations: $1.23\cdot 10^{-26}$}  \\
\hspace{10mm} $B$            &0.745                &$\pm$0.009    & \multicolumn{4}{c|}{~}  \\
\hspace{10mm} $B2$           &0.735                &$\pm$0.009    & \multicolumn{4}{c|}{Predicted sum of weighted residuals : 0.00639}   \\
\hspace{10mm} $V1$           &0.533                &$\pm$0.013    & \multicolumn{4}{c|}{Observed sum of weighted residuals : 0.00640}   \\
\hspace{10mm} $V$            &0.542                &$\pm$0.015    & \multicolumn{4}{c|}{~}   \\
\hspace{10mm} $G$            &0.463                &$\pm$0.012    & \multicolumn{4}{c|}{~}  \\ [2.ex]
\hline
\end{tabular}
\end{center}
\end{table*}

\begin{figure*}
\infig{17.0}{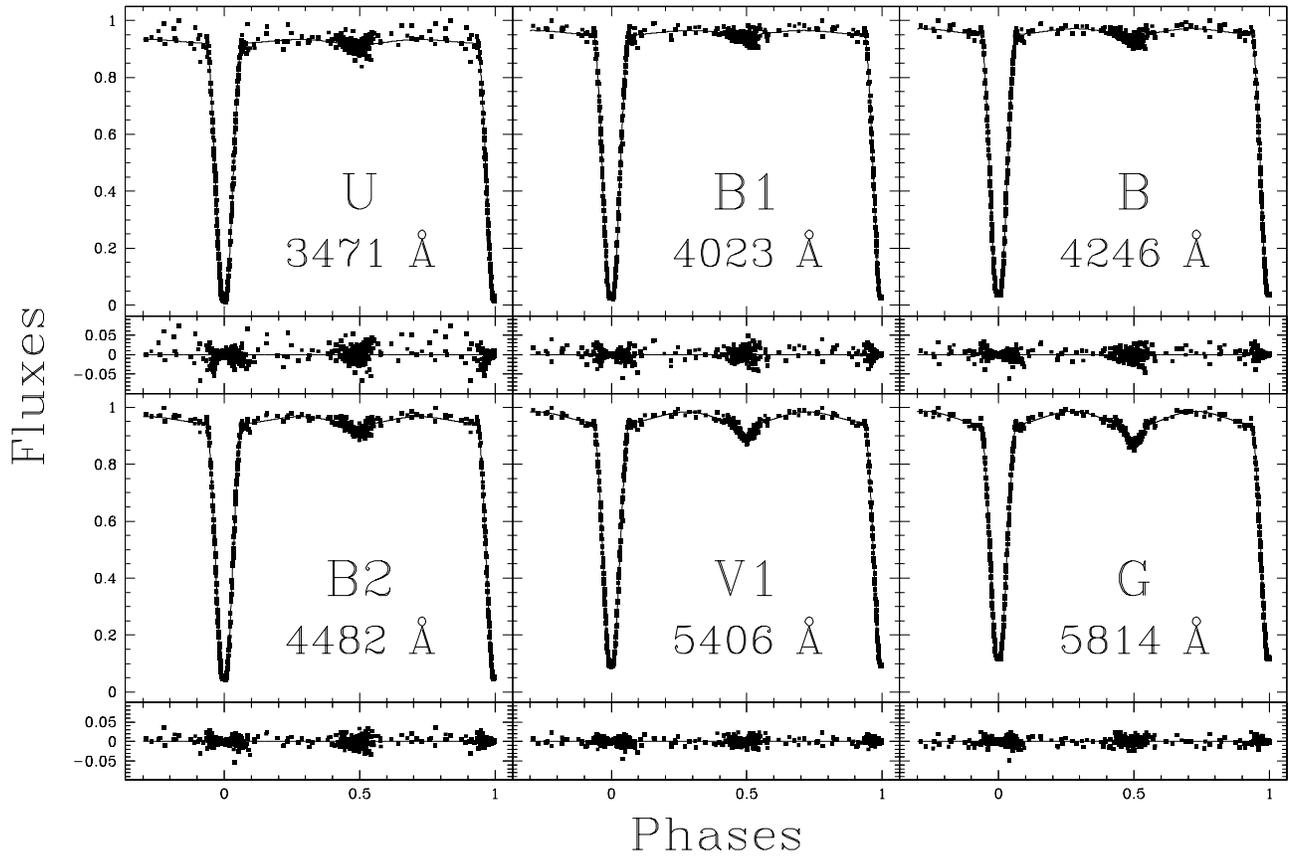}{17.0}
\caption[]{Light curves of TZ~Eri in 6 of the 7 Geneva photometric passbands 
($U$, $B1$, $B$, $B2$, $V1$, $G$). The light curve in V is given in Fig. 5.}
\end{figure*}
\begin{figure*}
\infig{17.0}{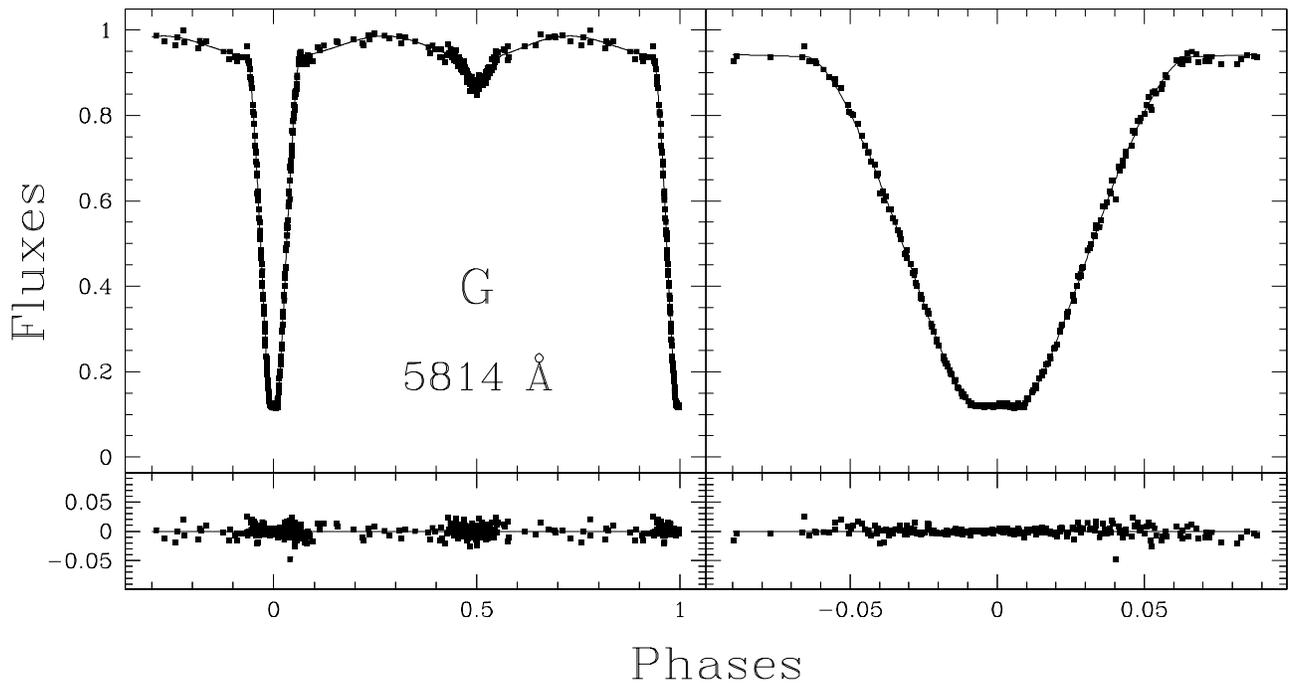}{17.0}
\caption[]{The light curve of TZ~Eri in magnitude V, with an enlargement of the primary eclipse.}
\end{figure*}

The photometric solution for TZ~Eri eclipses was obtained with the Wilson-Devinney WD programme
(Wilson \& Devinney, 1971; Wilson, 1992), using the version revised in 1995.
We used the WD programme in \textsc{Mode}~5, adapted for semi-detached 
systems , allowing a 
simultaneous computation on the seven Geneva photometric light curves, based on 393 measurements
(see Section~3), and the radial velocity curves of both components (see Table~4).

Lobe filling of the cool loser component was assumed. The semi-major 
axis of the relative orbit was initially set to $A = 10.2$ R$_{\odot}$ as calculated from radial
velocity results (see Section~5). Orbital eccentricity is fixed to zero and longitude of the 
periastron to $90 \degr$. Primary star temperature was fixed to 7770~K as determined
in Section~4. For both components, the stellar atmosphere models of Kurucz (1994) integrated by
Nicolet (1998) through the Geneva photometry passbands (Rufener \& Nicolet, 1988) have been used.

The bolometric albedos for hot and cool components were taken at the theoretical value of $1.0$ and
$0.5$ respectively (radiative and convective cases). Noise was set to 1 because scintillation can been neglected,
the photon noise being dominant (see Bartholdi et al., 1984). The grid resolution values were taken 
as 20, 20, 20, 20 for N1, N2, N1L and N2L 
respectively (see WD programme). For each of the seven Geneva magnitudes, the passband mean wavelength 
$\lambda_{0}$ was the one calculated by Rufener \& Nicolet (1988) as shown in Table 6. We assigned the
$\lambda_{0}$ value for the $B$ filter to the passband of the spectrographs used for the radial velocity
measurements. Stellar rotation is assumed to be synchronized for both components. 
For both the primary and secondary components, a logarithmic limb-darkening law of the form~:

\begin{equation}
I = I_0(1-x+x\cos\theta - y\cos\theta \ln(\cos\theta))
\end{equation}
was assumed (Van Hamme, 1993). For the secondary, both $x$ and $y$ parameters
were fixed to their theoretical values, interpolated from Table 2 of Van Hamme (1993).
Indeed, the secondary minimum is too shallow to allow the determination of $x_2$.
For the primary, the $y_1$ parameter was fixed to its theoretical value, because
it is not possible in practice to determine both $x_1$ and $y_1$ parameters since 
they are strongly correlated. The $x_1$ parameter, on the contrary, was left free and was fitted in the
least-squares procedure, for each of the seven passbands. The adopted values
of $y_1$, $x_2$ and $y_2$ are listed in Table 7, while the fitted values of
$x_1$ are given in Table 8.
The gravity darkening exponent $g_2$ is relevant to the secondary component and is
defined by the expression (Wilson \& Biermann, 1976)~:

\begin{equation}
T_{\rm local} = T_{\rm r.p.}\left(\frac{a_{\rm local}}{a_{\rm r.p.}}\right)^{0.25 g}
\end{equation}
where $T_{\rm local}$ is the local effective temperature, $T_{\rm r.p.}$
is the temperature at a reference point which here is the component's pole,
and $a_{\rm local}$, $a_{r.p.}$ are the accelerations due to gravity at
the same respective points.

The adjustable parameters are then semi-major axis $A$, inclination $i$, mass ratio $q=M_{2}/M_{1}$, cool star temperature $T_{2}$,
hot and cool star luminosities $L_{1}$ and $L_{2}$ in each passband, cool star gravity darkening 
exponent $g_{2}$ (see Wilson \& Biermann, 1976), hot star limb darkening coefficient $x_{1}$ in each passband
and potential $\Omega_{1}$ at the hot star surface. In reality, $\Omega$ is a non-dimentional parameter which
is a linear function of the true potential $\Psi$ (Kopal, 1959; Wilson \& Devinney, 1971). First, preliminary
light curves were generated using the {\it Light Curve}  sub-programme of WD until a satisfactory fit to the
observed light curves was obtained. Second, the method recommended by Van Hamme \& Wilson (1986)
was applied for the determination of the temperature and luminosity of the cool component. In a first step,
$T_{2}$ was determined with IPB option (see the WD programme) equal to 0, i.e. the luminosity is coupled to the temperature. 
In a second step, temperature and luminosity were de-coupled ($IPB = 1$), and it was possible to determine $L_{2}$. 
Finally, the following sets of parameters were adjusted~: first, successively ($T_{2}$, $\Omega_{1}$, $i$),
($L_{1}$, $L_{2}$), ($A$, $i$), ($x_{1}$, $g_{2}$), and then simultaneously ($A$, $i$, $\Omega_{1}$, $q$, $L_{1}$,
$L_{2}$), as recommended by the users of the WD programme. The solution was carried out to 
the point where the probable errors of all these parameters were smaller than the computed parameter corrections. 

Since the photometric temperature of the primary star is determined with an uncertainty of $\pm 100$ K (see Section~4), 
it is interesting to calculate the solution also for the two $T_{1}$ values 7670 K and 7870 K. The  
results are presented in Table 8. The differences are very small, except of course on $T_{2}$. 

It is especially
noteworthy that, for $T_{1}= 7770$ K, the value of $T_{2}$ ($4563 \pm 2$ K) is very close to the value obtained
in Section~4 on the basis of the photometric calibrations. The mass ratio $q=M_{2}/M_{1} = 0.1865 \pm 0.0003$ 
in Table~8 is a little smaller than the value derived from the radial velocity analysis ($K_{1}/K_{2} = 0.193 \pm 0.013$, see Table~5),
but still well within the uncertainty of the spectroscopic determination.
A similar difference can be noted on the semi-major axis of the relative orbit $A = a_{1}+a_{2}$ ($10.16 \pm 0.22$ R$_{\odot}$
in Table~5 and $10.57 \pm 0.16$ R$_{\odot}$ in Table~8). These differences are not surprising. We have to recall
that the radial velocity curve for the secondary component is based on only one measurement. The constraints imposed
by the photometric measurements of the eclipses produced an improvement of the orbital and physical parameters
of TZ~Eri. 

The small difference between the photometric and spectroscopic values of $q$ could indicate that the lobe 
filling by the secondary is not complete. If this were true, the real value of the uncertainty on $q$ would be larger than 
the photometric value (0.0003), may be as large as the spectroscopic value (0.013). Nevertheless, test solutions
have been obtained with the cool star slightly detached from its Roche lobe, and the results are less satisfactory
than those obtained in \textsc{Mode}~5 of the WD code~: as the cool star became farther detached from the lobe,
solution errors grew. This confirms the semi-detached status of the system.

The uncertainties on the derived parameters are of three types~:

-- Intrinsic, i.e. resulting from the mathematical analysis of the light and radial velocity curves 
(e.g. $\pm$ 2 K on $T_2$, see Table~8). Note that the intrinsic errors on $R_{1,2}$ for $A$ fixed are 
an order of magnitude smaller than the values in Table~9. 

-- Strongly correlated with the determination of $A$. For example, in Table~9, the errors on 
$M_{1,2}$, $R_{1,2}$ and $\log g_{1,2}$ almost entirely come from the uncertainty on $A$, and are 
pair wise strongly correlated (as well as with $A$). 

-- Depending on the photometric determination of $T_{\rm eff_1}$. For example, the values of 
$M_{\rm bol_{1,2}}$ strongly depend on the adopted value for $T_{\rm eff_1}$.

The light curves for the seven filters $U$, $B_{1}$, $B$, $B_{2}$, $V_{1}$, $V$ and $G$ are shown in Figs. 4 and 5. 
The quality of the fits is clearly extremely good. Figs. 6 and 7 present two views of TZ~Eri. Fig.~6 is a ``classical''
representation of the two components in the equatorial plane. The 3-dimentional representation of the potential
(Fig.~7) allows a better understanding of the meaning of the Lagrange points and of the path for the flow of the 
material from the secondary when it fills its Roche lobe. 

\begin{table}[hbt]
\caption[]{Computed parameters of the system TZ~Eri from the Wilson-Devinney programme. The parameters are,  
         for the hot primary ($1$) and the cool secondary ($2$) components~: the mass $M$, the mean radius $R$, 
	 the surface gravity $\log g$, the bolometric magnitude $M_{\rm bol}$ and the various radii (in units of
	 semi-major axis) of the deformed
	 components, i.e. {\it pole} (perpendicular to the orbital plane), {\it point} (in the direction of the 
	 other component), {\it side} (in the orbital plane, in the direction perpendicular to the direction
	 of the other component) and {\it back} (in the direction opposite to the other component). The uncertainty on
	 $M_{\rm bol}$ is calculated with an uncertainty on $T_{\rm eff}$ of $\pm$ 100 K.}
\begin{center}
\begin {tabular}{llll} 
\hline\rule{0pt}{3.5ex}%
Parameters                  &Values                     &Values                     &Values                    \\
                            &$(T_{1}=$			&$(T_{1}=$		    &$(T_{1}=$		       \\
			    &${\bf 7770} K)$            &${\bf 7670} K)$            &${\bf 7870} K)$            \\[2.ex]
\hline\rule{0pt}{3.5ex}%

$M_{1}$ [M$_{\odot}$]       &1.97 $\pm$ 0.06            &1.98                       &1.97 \\
$M_{2}$                     &0.37 $\pm$ 0.01            &0.37                       &0.37 \\[2.ex]
$R_{1}$ [R$_{\odot}$]       &1.69 $\pm$ 0.03            &1.69                       &1.68 \\
$R_{2}$                     &2.60 $\pm$ 0.04           &2.60                       &2.60 \\[2.ex]
$\log g_{1}$                &4.28 $\pm$ 0.03           &4.28                       &4.28 \\
$\log g_{2}$                &3.17 $\pm$ 0.03           &3.17                       &3.17 \\[2.ex]
$M_{\rm bol_1}$             &2.36 $\pm$ 0.09           &2.41                       &2.32 \\
$M_{\rm bol_2}$             &3.74 $\pm$ 0.13           &3.69                       &3.64 \\[2.ex]
$r_{\rm pole_1}$  ~[A]      &0.1592 $\pm$ 0.0003        &0.1595                     &0.1585 \\
$r_{\rm point_1}$           &0.1598 $\pm$ 0.0003        &0.1601                     &0.1591 \\
$r_{\rm side_1}$            &0.1596 $\pm$ 0.0003        &0.1599                     &0.1589 \\
$r_{\rm back_1}$            &0.1598 $\pm$ 0.0003        &0.1601                     &0.1591 \\[2.ex]
$r_{\rm pole_2}$  ~[A]      &0.2282 $\pm$ 0.0001        &0.2281                     &0.2283 \\
$r_{\rm point_2}$           &0.34~~~~   $\pm$ 0.01      &0.3349                     &0.3351 \\
$r_{\rm side_2}$            &0.2374 $\pm$ 0.0001        &0.2373                     &0.2374 \\
$r_{\rm back_2}$            &0.2696 $\pm$ 0.0001        &0.2695                     &0.2697 \\[2.ex]
\hline
\end{tabular}
\end{center}
\end{table}

\section{Discussion}

\subsection {Limb-darkening}
Our adjusted values of $x_1$ (see Table~8) compare relatively well with the theoretical ones,
in the sense that the difference hardly exceeds 0.1, or 15-20\% of the
value (except for $V$ where the difference reaches 0.15). Nevertheless, our 
values are systematically smaller than the
theoretical ones (computed for a solar chemical composition).
The discrepancies are larger than the formal errors
indicated in Table~8, which may partly, but probably not entirely be
explained by the fact that the errors listed should represent only lower
limits to the true ones.

Fig. 8 illustrates the dependance of $x_1$ on wavelength for a theoretical
star with $T_{\rm eff}= 7770$~K and $\log g = 4.28$, for the passbands
of Str\"omgren's $uvby$ system and Johnson's $UBV$ system (the values
have been linearly interpolated in Table 2 of Van Hamme, 1993). The fitted
values for the Geneva $U, B1, B, B2, V1, V, G$ passbands are shown
for comparison.

\begin{figure}
\infig{8.8}{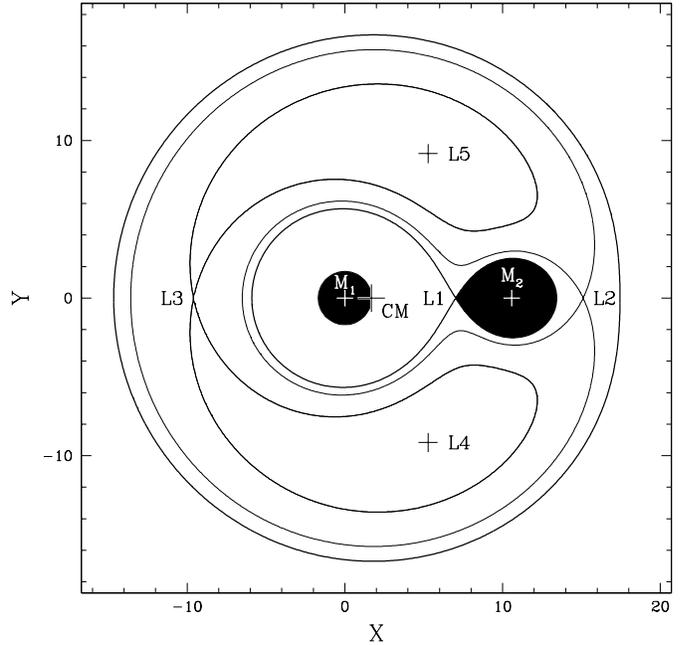}{8.8}
\caption[]{Schematic view of the system in the equatorial plane (coordinates 
in $R_{\odot}$). Some equipotential lines and the Lagrange points are indicated. 
The secondary star fills its Roche lobe.}
\end{figure}

\begin{figure}
\infig{8.8}{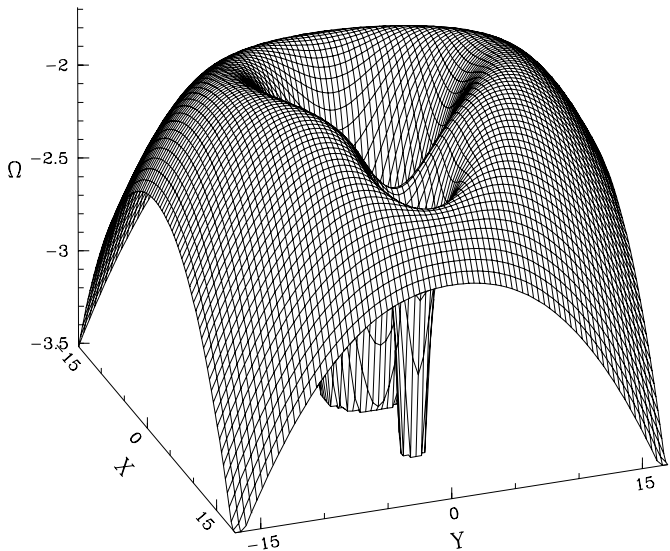}{8.8}
\caption[]{Schematic view of the gravity potential $\Omega$ (see Section~7 for the definition) 
of the TZ~Eri system. The horizontal coordinates are in $R_{\odot}$.}
\end{figure}

\begin{figure}
\infig{7.8}{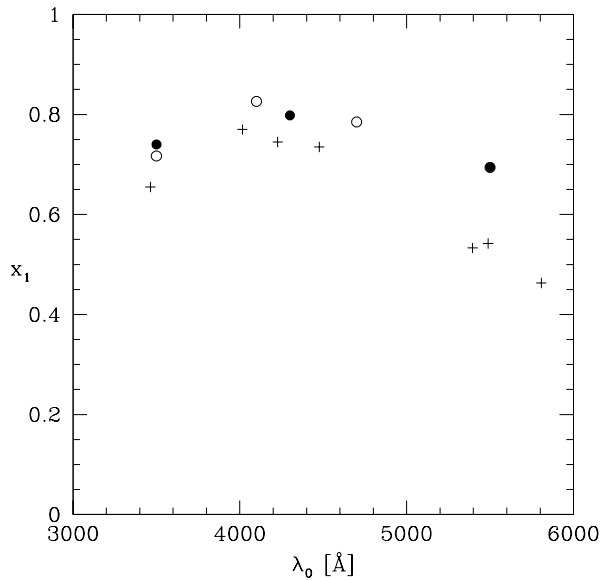}{7.8}
\caption[]{Limb-darkening $x_1$ coefficient of the primary as a function
of wavelength. Points are theoretical heterochromatic values for the $uvby$
(open dots) and $UBV$ passbands (full dots), while crosses are for empirical 
values determined in this work for the passbands of the Geneva system.
Notice that the theoretical points for the $v$ and the $V$ bands are
superimposed.}
\end{figure}

The theoretical dependance of $x_1$ on wavelength satisfies very well the
observations from the $U$ up to the $B2$ passband, except for a small,
systematic vertical shift. For the $V1$ to $G$ passbands, however, the
observed $x_1$ parameter is much smaller than the theoretical one. Since the
error bars are not larger than the symbols in Fig.~8 (except marginally for $U$), 
there is undoubtedly a significant discrepancy which remains to be explained. We
obtained a simultaneous solution in the seven passbands, thus at least the relative 
values of $x_1$ should be reliable. But, even considering only differential values 
(e.g. $x_1(V) - x_1(B)$) a difference with the theoretical predictions does remain. 
Unfortunately, we did not see any recent discussion in the literature about the
reliability of empirical limb-darkening coefficients in Algol-type systems.

\subsection {Distance}
From the values of $V$ (Table~2), $M_{\rm bol}$ (Table~9), $E[B2-V1]$ and $[B2-V1]_0$ (Section~4),
we derive a distance of $270 \pm 12$ pc for TZ~Eri, by adopting a zero value for the bolometric 
correction $BC$ of the primary, according to the $BC$-colour relation of Flower (1977).

\subsection {Algol-type stars}
Sarma et al. (1996) gave a detailed discussion on the evolutionary status of Algol 
components, on the basis of a comparison of their global parameters (masses, radii, luminosities, 
temperatures) with those of normal stars. As these authors noted~:

\begin{enumerate}
\item
   While primaries (mass gainers) of some semi-detached systems are overluminous,
   oversized and hotter for their masses, some others are underluminous, smaller and cooler.
\item
   Except for a few cases, the primaries are lying either on the main sequence or near it.
\item
   The secondaries (mass losers) have evolved off the main sequence (they were originally
   the more massive components).
\end{enumerate}

Figs. 9 and 10 present the Mass--Luminosity diagram ($\log L$ vs. $\log M$) and the HR diagram
($\log L$ vs. $\log T$) for the Algol systems in the lists by Sarma et al. (1996) and Maxted \&
Hilditch (1996) having a mass of the primary smaller than 4~M$_{\odot}$. The 3 systems with 
primaries off the main sequence (RZ~Cnc, AR~Mon, AW~Peg) are not represented. The underluminous
(black dots) and overluminous (black squares) have been separated in Fig. 10 by using the ZAMS 
(Schaller et al., 1992) to make the separation. We can see that~:

\begin{enumerate}
\item
   TZ~Eri belongs to the group of the underluminous primaries.
\item
   The mass of TZ~Eri's primary is the smallest of the group of these underluminous
   primaries.
\item
   The luminosity of TZ~Eri's primary is that of a main sequence star of 1.70 $M_{\odot}$
   located near the ZAMS (see Fig.~10), while the determined mass is 1.98 M$_{\odot}$ (see Table~9).
\item
   With respect to a main sequence star of the same mass, TZ~Eri's primary is underluminous by $\sim$0.3
   in $\log L$ (or fainter by $\sim$0.75 in $M_{\rm bol}$) and cooler by $\sim$900~K in $T_{\rm eff}$.
\item
   The mass transfer from the secondary to the primary has been important, taking into account the present
   masses of the components, i.e. 0.37 M$_{\odot}$ and 1.98 M$_{\odot}$. The minimum value of this transfered
   mass is given by~:
  
   $\Delta M = M_{1} - (M_{1} + M_{2})/2 \simeq 0.80 M_{\odot}$. 
\end{enumerate}

\begin{figure}
\infig{8.8}{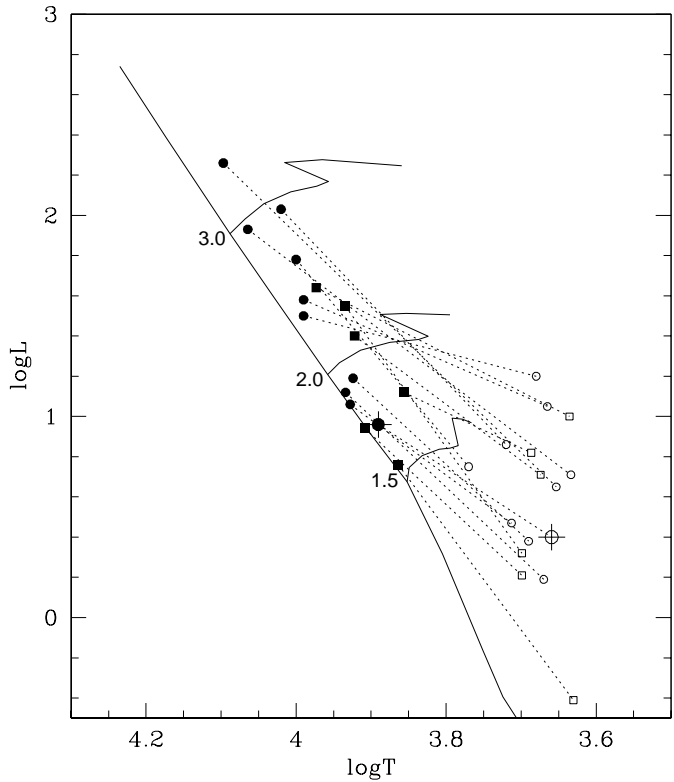}{8.8}
\caption[]{Theoretical HR diagram for the components of Algol systems with very precisely determined 
physical parameters according to the lists by Sarma et al. (1996) and Maxted \& Hilditch (1996), and 
to this paper for TZ~Eri (see Section~8.3). Primaries are identified with filled symbols and secondaries 
with open symbols. TZ~Eri is identified with plus symbols. Dots and squares refer to the systems having
respectively under- and overluminous primaries with respect to their masses (see Fig.~10). The Zero-Age 
Main Sequence and 3 evolutionary tracks (3.0, 2.0 and 1.5 $M_{\odot}$) are drawn, according to 
Schaller et al (1992).}
\end{figure}

\begin{figure}
\infig{8.8}{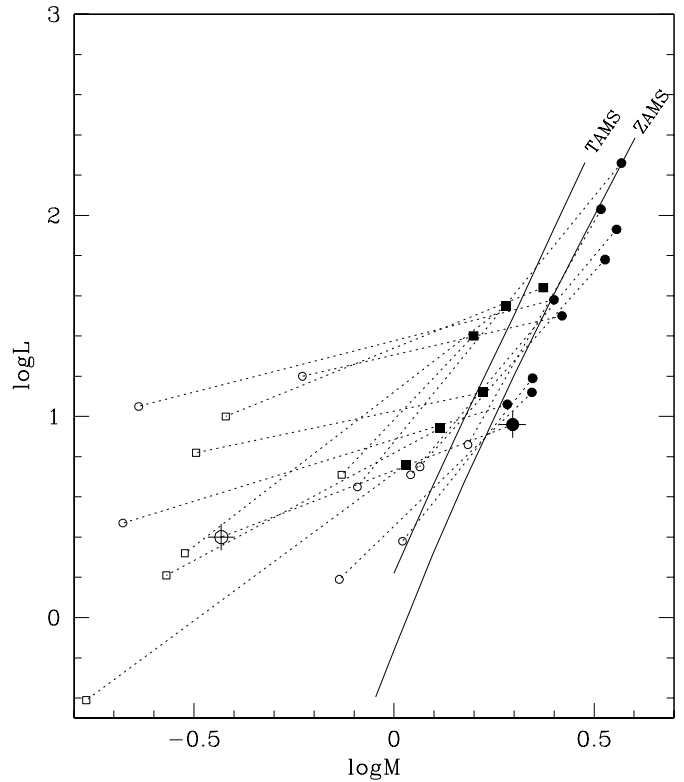}{8.8}
\caption[]{Mass-Luminosity relation for the components of the same Algol systems as those in Fig.~9. 
The zero-age main sequence (ZAMS) and the Terminal-Age Main Sequence (TAMS) are drawn, according to 
Schaller et al (1992). Primaries are identified with filled symbols and secondaries with open symbols. 
TZ~Eri is identified with plus symbols. Dots and squares refer to the systems having respectively 
under- and overluminous primaries with respect to the ZAMS.}
\end{figure}

\section {Conclusion}

The simultaneous adjustment of the light curve model on high precision photometric data in seven passbands, 
ranging from 3400 to 6000~\AA, put strong constraints on the physical and orbital parameters of TZ~Eri.
This is our first analysis of an Algol system on the basis of measurements made in the Geneva photometric
system. The essential results are presented in Figs.~4 and 5, Tables~8 and 9, and in the Abstract. The 
quality of the fits on the light curves and the uncertainties on the parameters show that this analysis
has been successful.

A further improvement on the determination of the physical parameters of the components of TZ~Eri, in
particular the mass ratio, would result from the acquisition of additional high S/N spectra of the
secondary cool component. Indeed, the present study is based on only one spectrum of the secondary and
this is not completely satisfying. However, this spectrum was obtained using the 3.5~m NTT telescope at
La~Silla and we were extremely lucky to have access to this large telescope for this single measurement.  

Accordind to Maxted \& Hilditch (1996), the number of Algol-type binary systems for which the absolute 
parameters are determined on the basis of a self-consistent solution of both the light and radial velocity 
curves is very limited. They mentioned only 9 systems for which masses, radii and luminosities are known
to accuracies typically better than 5\%. In this context, it is important to enlarge the sample of
very well known Algols, and  this was the aim of our study on TZ~Eri.

\acknowledgements{This work has been initiated by our late colleague Zdenek Kviz who obtained himself
                  the major part of the photometric data; indeed, we would have preferred to do this analysis with him. 
                  We are grateful to Drs. R.E.~Wilson for the programme of analysis of the light curves, 
		  B.~Nicolet for the treatment of the Kurucz's fluxes, D.~Raboud for the
		  NTT observations of the secondary, J.-C.~Mermilliod and J.~Andersen for the \textsc{Coravel} measurements,
		  M.~K\"unzli for the photometric observations of 1996 and M.~Studer for the reduction
		  of the NTT spectrum. All graphs have been produced using the SuperMongo package 
		  (Lupton \& Monger, 1998). This work has been partly supported by the Swiss National 
		  Science Foundation. The Mount Laguna Observatory is operated jointly by San Diego State 
		  University and the University of Illinois.}

\end{document}